\documentclass[journal, 10pt]{IEEEtran}

\usepackage{amsmath, amsthm, amsfonts, amssymb, amsbsy,nccmath}
\usepackage{mathtools}
\usepackage{bbm}
\usepackage{acronym}  
\usepackage[dvips]{color}
\usepackage{epsf}
\usepackage{times}
\usepackage{epsfig}
\usepackage{notoccite}
\usepackage{graphicx}
\usepackage{epstopdf}
\usepackage{pstricks}
\usepackage{amssymb}
\usepackage{amsxtra}
\usepackage{here}
\usepackage{rawfonts}
\usepackage{float}
\usepackage{times}
\usepackage{url}
\usepackage{cite}
\usepackage{array, makecell} %
\usepackage{caption}
\usepackage{subcaption}
\usepackage{algorithm}
\usepackage{algpseudocode}
\usepackage{blindtext}
\usepackage{enumitem}
\usepackage{xcolor, cite,etoolbox}
\usepackage{relsize}
\usepackage{lipsum}
\usepackage{graphicx}
\usepackage{tabularx}
\usepackage{xparse}
\usepackage{array}
\newcolumntype{P}[1]{>{\centering\arraybackslash}p{#1}}
\usepackage{mleftright}

\usepackage{mathtools}

\usepackage{graphics}
\usepackage{physics}
\usepackage{amssymb}
\usepackage{siunitx}
\include{newcommands}
\usepackage{multicol}

\usepackage[nomain,acronym,shortcuts]{glossaries}
\makeglossaries
\newcommand*{\acro}[3][]{\newacronym[#1]{#2}{#2}{#3}}

\acro{D2D}{device-to-device}
\acro{SC}{semantic communication}
\acro{SIR}{signal-to-interference-ratio}
\acro{SINR}{signal-to-interference-plus-noise-ratio}
\acro{PCP}{Poisson cluster process}
\acro{AD}{algorithm distillation}
\acro{NN}{neural network}
\acro{PRB}{physical resource block}
\acro{CoMP}{coordinated multi-point}
\acro{BS}{base station} 
\acro{MD-CoMP}{macrodiversity CoMP transmission}
\acro{MAC}{medium-access-control}
\acro{RAN}{radio access network}
\acro{JT-CoMP}{joint transmission CoMP}
\acro{CoMP-JT}{coordinated multipoint joint transmission}
\acro{SBS}{small base station}
\acro{MDSD}{multiple devices to a single device}
\acro{MDS}{maximum distance separable}
\acro{SCN}{small cell network}
\acro{PPP}{Poisson point process}
\acro{TCP}{Thomas cluster process}
\acro{CSI}{channel state information}
\acro{PDF}{probability distribution function}
\acro{PMF}{probability mass function}
\acro{RV}{random variable}
\acro{i.i.d.}{independently and identically distributed}
\acro{MBMS}{multimedia broadcasting multicasting service}
\acro{EE}{energy efficiency}
\acro{HCP}{hard-core placement}
\acro{CCDF}{complementary cumulative distribution function}
\acro{CDF}{cumulative distribution function}
\acro{PC}{probabilistic caching}
\acro{RC}{random caching}
\acro{CPF}{caching popular files} 
\acro{PGFL}{probability generating functional}
\acro{KKT}{Karush-Kuhn-Tucker}
\acro{PGF}{point generating function}
\acro{SCA}{successive convex approximation}
\acro{HD}{high-definition}
\acro{OSI}{open systems interconnection}
\acro{FHD}{full-high-definition}
\acro{UHD}{ultra-high-definition}
\acro{VR}{virtual reality}
\acro{AR}{augmented reality}
\acro{5G}{fifth-generation}
\acro{QoS}{quality-of-service}
\acro{QoE}{quality-of-experience}
\acro{IoT}{internet of things}
\acro{MHCPP}{Matern hardcore point process}
\acro{LoS}{line-of-sight}
\acro{NLoS}{non-line-of-sight}
\acro{PSD}{power spectral density}
\acro{MEC}{mobile edge computing}
\acro{E2C}{edge-to-cloud}
\acro{E2E}{end-to-end}
\acro{THz}{terahertz}
\acro{CLT}{central limit theorem}
\acro{HQ}{High Quality}
\acro{eMBB}{enhanced mobile broadband}
\acro{URLLC}{ultra reliable low latency communications}
\acro{mmWave}{millimeter wave}
\acro{EVT}{extreme value theory}
\acro{GEV}{generalized extreme value}
\acro{LIS}{large intelligent surface}
\acro{RIS}{reconfigurable intelligent surface}
\acro{RF}{radio frequency}
\acro{UE}{user equipment}
\acro{MIMO}{multiple-input multiple-output}
\acro{EVaR}{entropic value-at-risk}
\acro{DNN}{deep neural network}
\acro{MDP}{Markov decision process}
\acro{RL}{reinforcement learning}
\acro{RNN}{recurrent neural network}
\acro{GNN}{graph neural network}
\acro{ANN}{artificial NN}
\acro{LSTM}{long short-term memory}
\acro{ReLu}{rectified linear unit}
\acro{VaR}{value-at-risk}
\acro{SNR}{signal-to-noise ratio}
\acro{AoSA}{array of subarray}
\acro{XR}{extended reality}
\acro{AoA}{angle of arrival}
\acro{ULA}{uniform linear array}
\acro{AoD}{angle of departure}
\acro{EM}{electromagnetic}
\acro{HRLLC}{s high-rate and high-reliability low latency communications}
\acro{6DoF}{six degrees of freedom}
\acro{MR}{mixed reality}
\acro{PAPR}{peak to average power ratio}
\acro{OFDM}{orthogonal frequency-division multiplexing}
\acro{OFDMA}{orthogonal frequency-division multiple access}
\acro{SC-FDM}{single carrier frequency-division multiplexing}
\acro{ToA}{time of arrival}
\acro{MUSIC}{multiple signal classification}
\acro{IoE}{Internet of Everything}
\acro{DT}{digital twin}
\acro{PT}{physical twin}
\acro{CT}{cyber twin}
\acro{DRL}{deep reinforcement learning}
\acro{FL}{federated learning}
\acro{DL}{deep learning}
\acro{CRAS}{connected robotics and autonomous system}
\acro{CL}{continual learning}
\acro{BF}{beamforming}
\acro{MSE}{mean squared error}
\acro{EWC}{elastic weight consolidation}
\acro{ML}{machine learning}
\acro{GD}{gradient descent}
\acro{MLP}{multi layer perceptron}
\acro{TL}{transfer learning}
\acro{AI}{artificial intelligence}
\acro{NFT}{non fungible token}
\acro{H2A}{human-to-avatar}
\acro{A2A}{avatar-to-avatar}
\acro{UAV}{unmanned aerial vehicle}
\acro{NTN}{non-terrestrial network}
\acro{ISAC}{integrated sensing and communication }
\acro{CIS}{connected intelligence system}
\acro{QoVE}{quality of virtual experience}
\acro{OOD}{out-of-distribution}
\acro{XAI}{explainable AI}
\acro{LLM}{large language model}
\acro{CNN}{convolutional neural network}
\acro{KPI}{key performance indicator}
\acro{SCM}{structural causal model}
\acro{CGM}{causal graphical model}
\acro{DAG}{directed acyclic graph}
\acro{IIT}{integrated information theory}
\acro{CI}{connected intelligence}
\acro{UL}{uplink}
\acro{KF}{Kalman filter}
\acro{GAN}{generative adversarial networks}
\acro{6G}{sixth generation}
\acro{MCMC}{Markov chain Monte-Carlo}
\acro{MAB}{multi-armed bandit}

\usepackage{datetime}
\usepackage{amssymb}
\usepackage{subcaption}
\usepackage{tcolorbox}
\usepackage{booktabs}

\usepackage{verbatim}

\newcommand{\beq}{\begin{equation}}
\newcommand{\eeq}{\end{equation}}

\def\adots{\mathinner{\mskip0mu\raise0pt\vbox{\kern7pt\hbox{.}}\mskip3mu
          \raise4pt\hbox{.}\mskip3mu\raise8pt\hbox{.}\mskip0mu}}

\usepackage{bm}

\newcommand{\bms}{{\bm s}}

\newcommand{\bma}{{\bm a}}

\begin{document}
\title{Large Multi-Modal Models (LMMs) as Universal Foundation Models for AI-Native Wireless Systems \vspace{-0.7cm}
\author{Shengzhe~Xu, 
Christo Kurisummoottil Thomas,~\IEEEmembership{Member, IEEE},
Omar Hashash,~\IEEEmembership{Graduate~Student~Member,~IEEE}, 
Nikhil~Muralidhar,
Walid Saad, \IEEEmembership{Fellow, IEEE}, 
and~Naren Ramakrishnan, \IEEEmembership{Fellow, IEEE}\vspace{-0.5cm}}%
\thanks{S. Xu and N. Ramakrishnan are with the Sanghani Center for Artificial Intelligence and Data Analytics at Computer Science Department of Virginia Tech. Emails: shengzx@vt.edu, naren@cs.vt.edu.} 
\thanks{C. K. Thomas, O. Hashash, and W. Saad are with the Bradley Department of Electrical and Computer Engineering, Virginia Tech, Arlington, VA, USA. Emails: \protect{christokt@vt.edu}, \protect{omarnh@vt.edu}, \protect{walids@vt.edu}.\\
  \indent  N. Muralidhar is with the Department of Computer Science at Stevens Institute of Technology, NJ, USA. Email: nmurali1@stevens.edu.}
}
\maketitle
\begin{abstract}

Large language models (LLMs) and foundation models have been recently touted as  a game-changer for 6G systems. However,  recent efforts on LLMs for wireless networks are limited to a direct application of existing language models that were designed  for natural language processing (NLP) applications. 
To address this challenge and create wireless-centric foundation models, this paper presents a comprehensive vision on how to design \emph{universal foundation models} that are tailored towards the unique needs of  next-generation wireless systems, thereby paving the way towards the deployment of \emph{artificial intelligence (AI)-native networks}. Diverging from NLP-based foundation models, the proposed framework promotes the design of \emph{large multi-modal models (LMMs)} fostered by three key capabilities: 1) processing of \emph{multi-modal sensing} data, 2) \emph{grounding} of physical symbol representations in real-world wireless systems using causal reasoning and retrieval-augmented generation (RAG), and 3) enabling \emph{instructibility} from the wireless environment feedback to facilitate dynamic network adaptation thanks to logical and mathematical reasoning facilitated by neuro-symbolic AI. In essence, these properties enable the proposed LMM framework to build universal capabilities that cater to various cross-layer networking tasks and \emph{alignment} of intents across different domains. 
Preliminary results from experimental evaluation demonstrate the efficacy of grounding using RAG in LMMs, and showcase the alignment of LMMs with wireless system designs. 
Furthermore, the enhanced rationale exhibited in the responses to mathematical questions by LMMs, compared to vanilla LLMs, demonstrates the logical and mathematical reasoning capabilities inherent in LMMs. Building on those results, we present a sequel of open questions and challenges for LMMs. We then conclude with a set of recommendations that ignite the path towards LMM-empowered AI-native systems.
%
\end{abstract}
\begin{IEEEkeywords}\vspace{-0mm}
Large multi-modal models, Universal foundation model, AI-native, Grounding, Alignment, Instructibility.
\end{IEEEkeywords}

\section{Introduction}

Future \ac{AI}-native wireless systems (e.g., 6G and beyond)  must leverage
\ac{ML} and \ac{AI} algorithms to design, optimize, and operate various facets of the network, including
resource allocation, interference management, transceiver design, and others~\cite{ChenComST2019}. 
Consequently, the cross-layer network functionalities implemented by \ac{AI} models could propel advanced network capabilities, that include: 1) \emph{resilience}, enabling 6G networks to withstand disruptions and maintain connectivity even in challenging scenarios; 2) \emph{intent management}, allowing networks to autonomously translate high-level business intents into closed-loop network configurations; 3)
\emph{big-data analytics}, enabling diagnostics using historical wireless data,  addressing software or hardware failures,  improving communication and computing resource usage, and predicting future user and network behavior; and 4) \emph{non-linear signal processing}, allowing networks to  process multi-modal signal characteristics, and propagation environments.  
To achieve the above goals, a promising avenue is to explore generative AI's universal knowledge retrieval and generation capabilities, particularly foundation models such as large language models (LLMs). Trained on diverse datasets, LLMs can discern intricate patterns and offer insights for optimizing end-user experience in future connected intelligence applications. In particular, LLMs excel in \emph{predictive analytics}, foreseeing potential network performance deviations based on historical data and enabling proactive measures. Moreover, LLMs facilitate \emph{natural language interactions} for network management, enhancing accessibility. Additionally, they could aid in optimizing signaling schemes, and resource allocation by recommending adaptive strategies, improving efficiency. 
In this context, the dynamic reconfigurability of these systems, facilitated by LLMs, could help expedite the incorporation of standardization into the system design across \ac{OSI} layers of future AI-native wireless systems. This, in turn, enables a faster time to market for the next-generation wireless products.

\begin{figure*}[t]
\centerline{\includegraphics[width=0.9\textwidth]{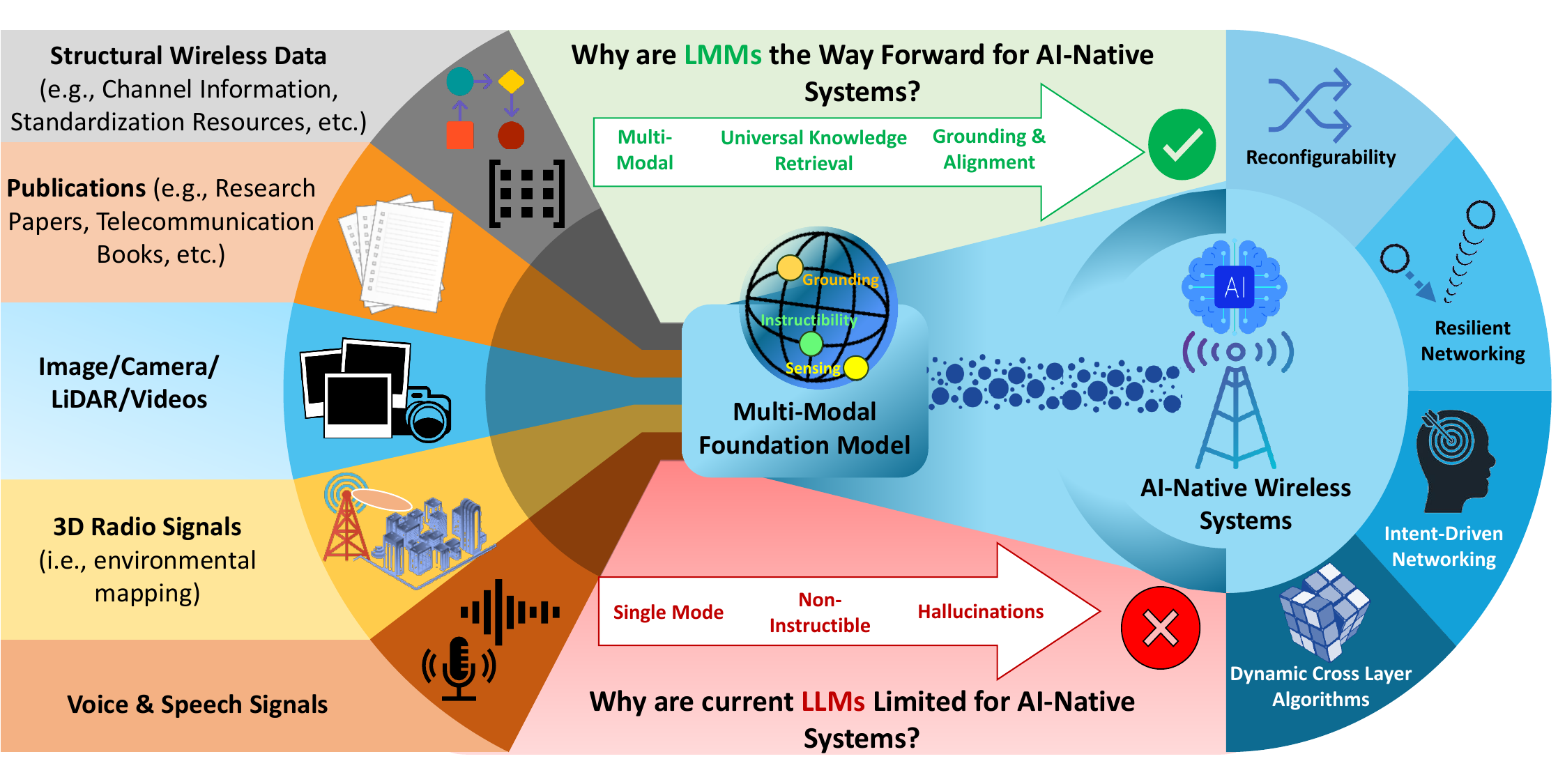}}\vspace{-1mm}
\caption{\small Illustrative figure of the proposed framework for LMM-empowered AI-native wireless systems.}
\label{CausalAI_Vision}\vspace{-0mm}
\vspace{-4.5mm}
\end{figure*}

\vspace{-1.5mm}\subsection{Related Works and Limitations}

LLMs for wireless networks have been recently explored in~\cite{MaatoukArxiv2023, TarkomaArxiv2023,HuangArxiv2023,BariahZhaoArxiv2023}. 
However, the LLMs considered in~\cite{MaatoukArxiv2023, TarkomaArxiv2023,HuangArxiv2023} are confined to processing a single mode of textual data, which restricts their role exclusively to network chatbots. Accordingly, such LLMs cannot capture the modality of the data rising from the multi-service functionalities (e.g., sensing, communication, etc.) of future wireless networks.
Although the authors in~\cite{BariahZhaoArxiv2023} present a vision focused on utilizing \emph{multi-modal} LLMs, their approach relies on LLMs like GPT-x, LLaMA, or Falcon tailored for natural language processing (NLP) tasks. To become effective, such multi-modal LLMs must be fine-tuned to single, narrow wireless tasks such as resource allocation. Consequently, fine-tuning limits their adoption as a \emph{universal} solution to different interrelated, cross-layer tasks in AI-native networks.

Furthermore, we identify the limitations of~\cite{MaatoukArxiv2023, TarkomaArxiv2023,HuangArxiv2023,BariahZhaoArxiv2023} that can hinder adopting LLMs as a transformative solution for meeting the AI-native network goals. \emph{First}, current works overlook how \ac{AI}-native networks can fuse, at scale, the environmental \emph{sensing} data that drive their multi-modal LLMs. Moreover, state-of-art solutions like~\cite{BariahZhaoArxiv2023} neglect the fact that even textual wireless data can be structured in specific formats such as tables (e.g., 3GPP tables) and network performance evaluations can be presented as in the form of graphs or images. The limitation in real-world sensing, in terms of structure and modality, restricts the range of functionalities (e.g., wireless chatbots, text to image conversion) for which LLMs can be effectively adopted.

\emph{Second}, current LLMs lack the essential \emph{grounding} abilities that connect their abstract, language-based knowledge to tangible real-world experiences. In fact, LLMs majorly gain their knowledge upon being trained on extensive corpuses of text data. Hence, these LLMs are unaware of the complex physics governing the wireless environment, such as the propagation process of wireless signals, thereby leading to potentially inconsistent decisions and predictions. The absence of grounding impedes AI-native networks from carrying out logical, causal, and mathematical reasoning operations, which is a cornerstone for achieving future network goals such as resilience, intent management, non-linear signal processing, and others. 
Therefore, it is necessary to ensure that the representations acquired by LLMs accurately interpret information from the real world and adhere to the network goals. 


\emph{Third}, a persistent challenge of LLMs is their tendency to hallucinate by generating ``human-like" outputs that do not connect to reality, essentially fabricating false information\cite{RawteArxiv2023}. If such hallucinations occur, AI-native networks driven by LLMs may generate misleading or inaccurate information. For example, the network may propose power allocations that may violate the regulated thresholds of transmit powers of base stations. Thus, relying on vanilla LLMs can lead to alignment problems in AI-native networks. 
\emph{Alignment} here can encompass fulfilling network objectives, adhering to physical constraints like radiated power or environmental sustainability goals, and ensuring compliance with governmental regulations that oversee the deployment of wireless networks. Hence, current LLMs lack precise \emph{instructibility} from the environment. This shortcoming impedes their ability to remain under control and guidance. Instructibility refers to the capability of LLMs to dynamically adjust their behavior based on explicit feedback from users, whether they be user equipments, system engineers, or network operators. This adaptability should also ensure that decisions made by LLMs are explainable, fostering trustworthy solutions.

Therefore, it is imperative to design \emph{multi-modal LLMs} endowed with \emph{sensing, grounding}, and \emph{instructibility} properties in an attempt to enable AI-native networks that can have \emph{alignment} abilities as well as \emph{universal} capabilities. Here, \emph{universal foundation model} for wireless systems can be defined as AI models tailored to handle a wide array of tasks and applications within the wireless domain, irrespective of the network architecture and standards. Furthermore, they must demonstrate proficiency in understanding the real-world workings of the wireless environment behind the sensed data. This includes the ability to reason and provide conclusions for novel scenarios and generate novel information across diverse facets of wireless communication, covering areas such as network management, resource allocation, protocol adherence, and troubleshooting.

\begin{figure*}[t]
\centerline{\includegraphics[width=0.75\textwidth]{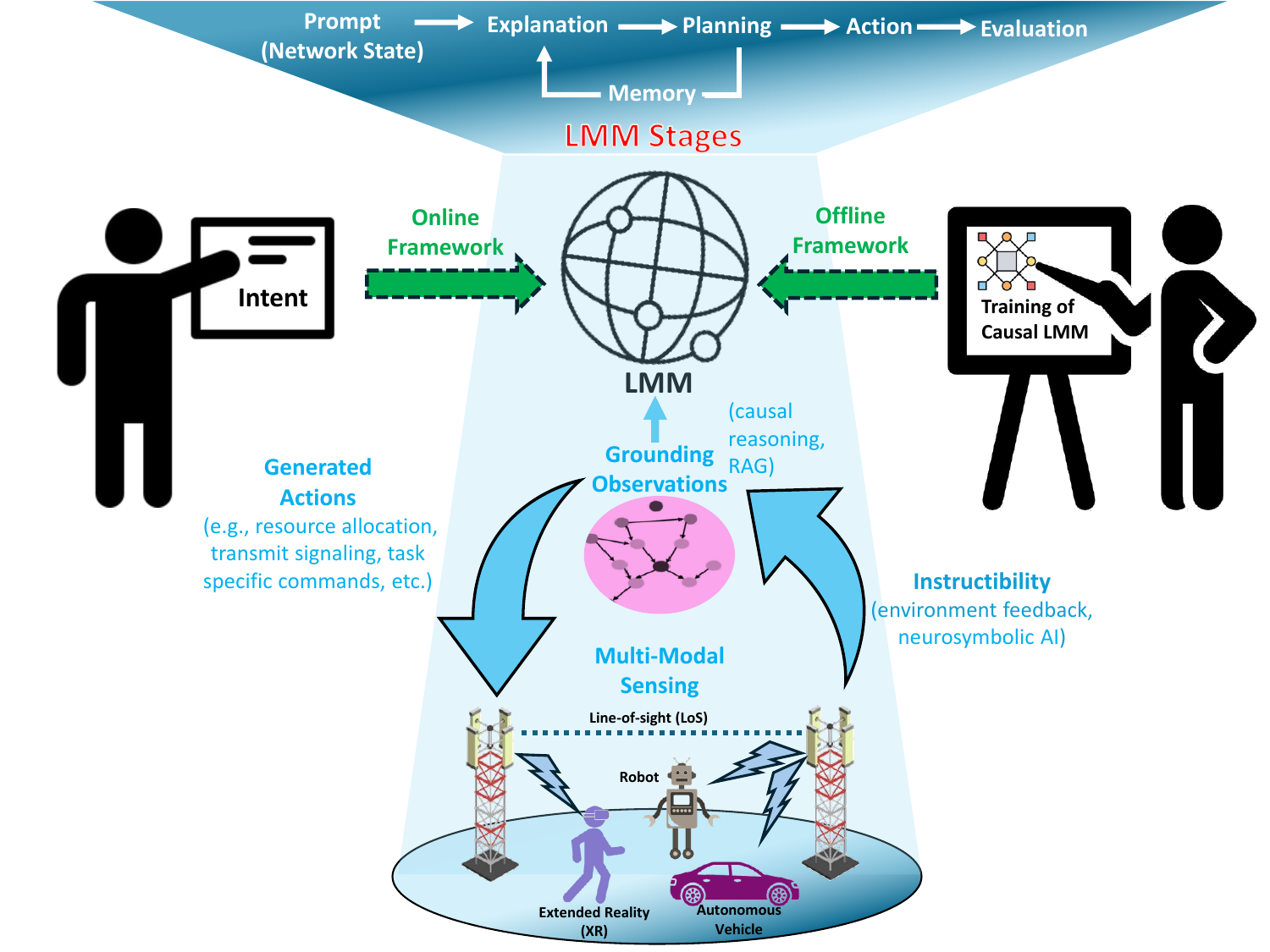}}\vspace{-1mm}
\caption{\small Proposed framework of wireless-centric LMMs  with capabilities of grounding, instructibility, and alignment. }
\label{Resilient_network_robot}\vspace{-0mm}
\vspace{-3.5mm}
\end{figure*}

\subsection{Contributions}
The main contribution of this paper is the introduction of of \emph{grounded and instructible} \emph{large multi-modal models (LMMs)} that are \emph{universal} and ensure that the AI-native wireless systems achieves \emph{alignment} capabilities, as shown in Fig.~\ref{CausalAI_Vision}. 
Our key contributions include:



\begin{itemize}
\item We first propose a novel framework for universal, wireless-centric foundation models, that goes beyond \cite{MaatoukArxiv2023, TarkomaArxiv2023, HuangArxiv2023} by integrating the following capabilities into LMMs: 1) \emph{Multi-modal data fusion}: fusing multi-modal sensing information to a shared semantic space thus enabling efficient training of universal foundation models, 2) \emph{Grounding:} involving the creation of a wireless-specific language through RAG~\cite{CartaArxiv2023} and leveraging causal reasoning, and 3) \emph{Instructibility:} facilitating transparent interactions between the wireless environment and LMMs through online \ac{RL} and interactive prompting and neuro-symbolic AI to perform logical and mathematical reasoning. This approach ensures \emph{alignment} through the development of trustworthy LMMs that can explain the reasons behind wireless data, propose cross-layer network actions aligned with network goals, and accommodate various physical constraints. 
\item We showcase the effectiveness of embedding wireless concepts into LMMs using RAG. Proof-of-concept results demonstrate that infusing more wireless context improves the accuracy of LMM responses, thereby reducing hallucinations compared to responses generated without wireless context. Furthermore, for mathematical questions, an LMM delivers accurate responses with proper rationale, showcasing its ability to reason effectively when grounded in the right context.
\item 

We present a use case of intent-based management employing LMMs, comprising problem formulation, intent assurance, and a validation phase. The improved logical and mathematical reasoning capabilities (shown in experiments) empower LMMs to function as dynamic problem solvers. The accurate responses generated by LMMs facilitate the validation of proposed solutions during intent management, ensuring alignment with any physical constraints.  
We also demonstrate that logical and mathematical reasoning capabilities enable continuous monitoring of network performance--a critical aspect of building resilient wireless networks. In contrast to deep \ac{RL}-based methods, which may be constrained to specific domains, LMMs can accelerate network service recovery during failures by proposing a sequence of remediation actions. This holds true for failures in software, hardware, or the wireless environment. 
\item 
We highlight challenges in constructing universal foundation models, covering aspects such as network planning, acquiring diverse datasets, and adapting to evolving standards. Additionally, we propose approaches to ease computational complexity in training and inference for LMMs, contributing towards sustainable wireless network goals. This involves leveraging model sparsity, context windowing, and building distributed foundation models.

\end{itemize}

In a nutshell, our comprehensive framework and the initial experimental results serve as a valuable guide for understanding the essential components necessary for building universal wireless foundation models that can effectively execute diverse cross-layer network functionalities.

\section{LMM-Empowered AI-Native Wireless Systems: Proposed Framework}

\label{Vision}

To build universal foundation models, as shown in Fig.~\ref{Resilient_network_robot}, the proposed multi-modal LMM framework is built upon the principles of 1) multi-modal data fusion, identifying the physical symbols present in the wireless environment and fusing the multi-modal data to a shared semantic space with minimal dimensions, and also avoiding data redundancy across multiple modalities.; 2) grounding, achieved through a blend of causal reasoning, and RAG, and which allows an understanding of the physical symbol meaning and the relations among them from wireless perspective; and 3) instructibility, enabling dynamic adaptation of signaling and resource allocation strategies based on environmental feedback, facilitated by logical and mathematical reasoning via neuro-symbolic AI. The resulting LMM-generated network actions align accurately with 3GPP standards and government regulatory norms. 
\begin{figure*}
    \centering
    \includegraphics[width=\linewidth]{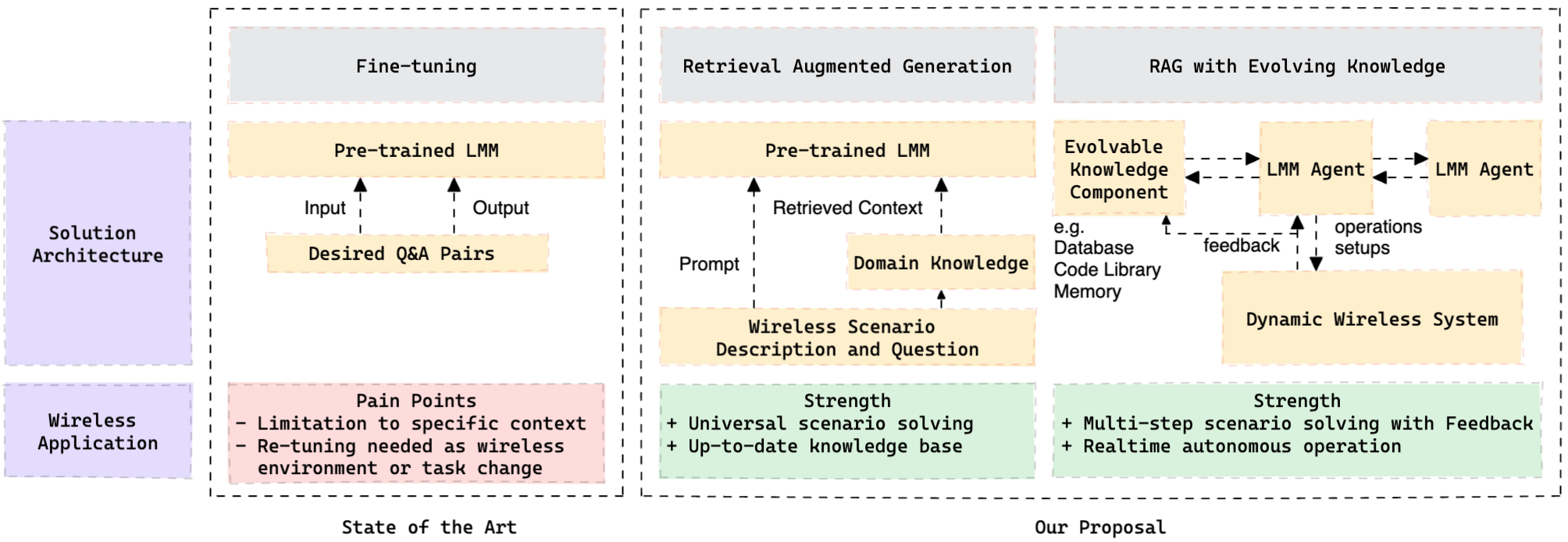}
    \caption{\small Applying LMM solutions for wireless applications: 1) Fine-tuning, 2) RAG, and 3) RAG with evolving knowledge.}
    \label{fig:llm_primer}\vspace{-3mm}
\end{figure*}
The components of the proposed universal foundation models framework are discussed next.

\vspace{-2mm}\subsection{Fusion of Multi-Modal Sensing Information: A Tradeoff Between Minimality and Redundancy}

The gateway to capturing the real-world wireless environment starts with the precise sensing and mapping of its diverse surroundings. In fact, previous works like \cite{BariahZhaoArxiv2023} (and references therein) discussed exploiting visual generative AI models such as DALL-E3 or multi-modal transformers such as meta-transformer to map multi-modal wireless sensing information to a semantic latent space. By capturing the characteristics of the wireless environment, such models can be used for enhancing contextual and situational awareness for sensing applications in AI-native networks. 
However, providing the entire mapped sensing information (received from diverse sources) as input to train the LMM is resource-intensive and requires substantial time for retraining. This hinders the timely execution of dynamic updates essential for maintaining seamless connectivity. 
To address this limitation of meta-transformers inherent in \cite{BariahZhaoArxiv2023}, we propose the use of data filtering. As discussed next, this data filtering approach, aims to convey a condensed sequence of pertinent information to the LMM.
To achieve this, we propose to start with the identification of physical symbols present across the multi-modal data. \emph{Physical symbols} here refer to abstract entities present in the data that have relevant semantics with respect to the wireless network. For example, this may involve associating symbols with various extrinsic and intrinsic elements. \emph{Extrinsic elements} encompass dynamic objects, such as scattering elements in the environment and users. Meanwhile, \emph{intrinsic elements} involve static network features like network addresses and signal processing methods underlying wireless transmission or reception, among others. 
Here, there can be redundancy among the information conveyed by the symbols across multiple modalities of data. To mitigate redundancy, the shared semantic latent space's construction should adhere to the information bottleneck principle \cite{anonymous2024neuroinspired}. In this framework, it is assumed that a dominant mode of information, called \emph{prime modality}, exists within a dataset, serving as the primary source of information. Other modalities complement or enhance the information provided by this prime modality in a multi-modal dataset. Here, the compact representation for any modality should convey as little information as possible about the raw data, and, simultaneously, prime modality representation should convey maximum information about other modalities. This ensures that the resulting semantic latent space is of \emph{minimal dimension} and at the same time \emph{avoids redundant information}. 
Further, for LMM training, we advocate using this filtered representations in the shared semantic space as inputs. 
Additionally, filtering also determines when to perform dynamic updates of the \ac{NN} parameters of the LMM, taking into account the nature of the captured data, which can be either static (e.g., 3GPP standards) or dynamic (e.g., wireless channel information or changing applications).

We have discussed the fusion and filtering of multi-modal information and the training of LMMs. 
However, on its own, simply identifying symbols is not enough if LMMs are to be employed for constructing universal foundation models capable of performing various cross-layer network functionalities. While a vanilla LLM may excel in statistically predicting the events following an observed sequence of sensing information, it lacks an understanding of what a specific observed wireless environment implies from a wireless networking perspective. For example, translating images of trees in a wireless environment into a set of angles of arrival or departure that describe the RF signal propagation environment requires associating meaning from a wireless perspective with each extracted physical symbol. This aspect, known as grounding, is detailed in the subsequent discussion. 

\vspace{-2mm}\subsection{Causal Reasoning for Grounding in LMMs: Reducing Hallucinations and Bolstering Trustworthiness}
\label{Grounding_Symbols}

As discussed, the challenge of grounding hinders the effectiveness of LLMs in real-world inference. 
Traditional grounding approaches typically entail the creation of a knowledge base, which represents an instance of symbolic AI. This method builds a complex conceptual network, functioning as the system's memory and directly aligning with real-world operations. The conceptual web captures the possible logical relations among physical symbols such as scattering objects, users, network topology, transmission or reception parameters, among others. 
Nevertheless, the use of knowledge base methods faces scalability challenges as the number of relations and physical symbols expands. To overcome this limitations in conventional grounding methods, we propose that LMMs infer the relations among various physical symbols identified using \emph{causal reasoning}, which is discussed below \cite{thomas2023causal}. 





While specific experiments demonstrate that language models might exhibit causality, it is predominantly attributed to the causal knowledge ingrained in the training data, rather than indicative of LLMs possessing inherent causal understanding. 
In \cite{ZhangArxiv2023}, a gradient-based, transformer-type algorithm for zero-shot optimal covariate balancing for causal treatment effect is introduced. We propose to advance \cite{ZhangArxiv2023} by incorporating theoretical methods to construct causal foundation models, focusing on wireless concepts as the relevant physical symbols. Here, one may ask: \emph{how to identify the causal relations among physical symbols and how to ensure that the learned relations are aligned with the wireless concepts in standards and textbooks?} 
A standard approach is to perform fine-tuning \cite{BariahZhaoArxiv2023} that takes a pre-trained language model trained on large amounts of general text and then continue to train it on a small-scale task-specific text. The fine-tuned \ac{NN} layers can capture the causal relations among wireless concepts and observations. Fine-tuning is appropriate if the user specifically knows the ground-truth causal relations. For wireless scenarios, fine-tuning can be beneficial for constructing a wireless specific chatbot capable of extracting valuable information from its knowledge base. However, fine-tuning LLMs may impose limitations on the wireless applications supported, confining them to specific contexts. This limitation arises from the narrow set of \ac{NN} parameters that are tuned during the fine-tuning process (limited degrees of freedom), necessitating re-tuning as the wireless environment or task undergoes changes. To address the above limitations, we suggest leveraging the capabilities of RAG, and then perform causal discovery as discussed next (also see Fig.~\ref{fig:llm_primer}).

\subsubsection{Causal discovery through RAG}
\label{RAG_fine-tuning}
\begin{itemize}

\item \textbf{RAG:} Different from fine-tuning, which supervises both input and output, RAG~\cite{lewis2020retrieval}
is designed to strengthen the input of prompting. Through querying from a domain-specific database, RAG bridges the gap between the corresponding domain-knowledge context and the original question in the prompt.
In wireless scenarios, the domain-knowledge database can be a textbook, 3GPP standard, or any device instruction handbook. Once this information is retrieved, the generation component of RAG can be employed to formulate new content that infers or expresses causal relations among various physical symbols identified. 
For example, when the LMM is tasked with deducing causal relationships between scattering objects in the environment and channel parameters like angle of arrival (AoA) or angle of departure (AoD), RAG facilitates the initial learning of wireless concepts from textbooks or standards. Subsequently, RAG assists in inferring causal relations among the observed wireless events.
\item \textbf{RAG with evolving knowledge:}
Rather than depending solely on LMMs, RAG with evolving knowledge~\cite{wang2023voyager} operates as a dynamic multi-step system with a feedback loop. Through an evolvable external knowledge component and multi-agent cooperation, RAG becomes versatile in implementing more demanding applications, requiring multi-tasking like in connected home or industrial robots, universal knowledge retrieval for semantic communications \cite{ChristoTWCArxiv2022}, and intent management. With its continuous learning capability, RAG with evolving knowledge enables continually updating the wireless algorithms across all \ac{OSI} layers, coupled with the advancements in semiconductor industry and sofware solutions. This proves advantageous, especially in the context of intent management and resilience (see Section~\ref{Intent_Management} and ~\ref{Resilience}). Apart from the continual learning capability, RAG with evolving knowledge enables the retrieved knowledge from either a textbook or the standard also to be dynamically adjusted to cater to the specific application demands. 
For instance, this means that in the case of a multi-user communication system, there is a need to retrieve literature on signal processing algorithms that are different from what might be required in a single-user scenario.
\end{itemize}

\subsubsection{What does causal discovery through RAG entail for LMMs?}

Grounding via causal discovery entails endowing LMMs with the capability to comprehend the causal relationships among physical symbols and subsequently engage in causal inference through interventions and counterfactuals \cite{thomas2023causal}. Through interventions and counterfactuals, the LMM can indulge in chain-of-thought kind of reasoning, where it analyzes a sequence of causal state-action pairs $(\bms_t,\bma_t)$ and their effects, $\bms_0 \xrightarrow[]{\bma_0} 
 \bms_1 \xrightarrow[]{\bma_1}  \bms_2 \cdots \xrightarrow[]{\bma_{N-1}}  \bms_{N}$. This ability facilitates long-term planning for wireless resource allocation, signaling schemes for transmission and reception (and may include beamforming, modulation, coding, and control signaling, among others), and \ac{QoS} management, thereby contributing to the establishment of robust and resilient wireless systems.

\begin{figure*}[t]
\centerline{\includegraphics[width=0.85\textwidth]
{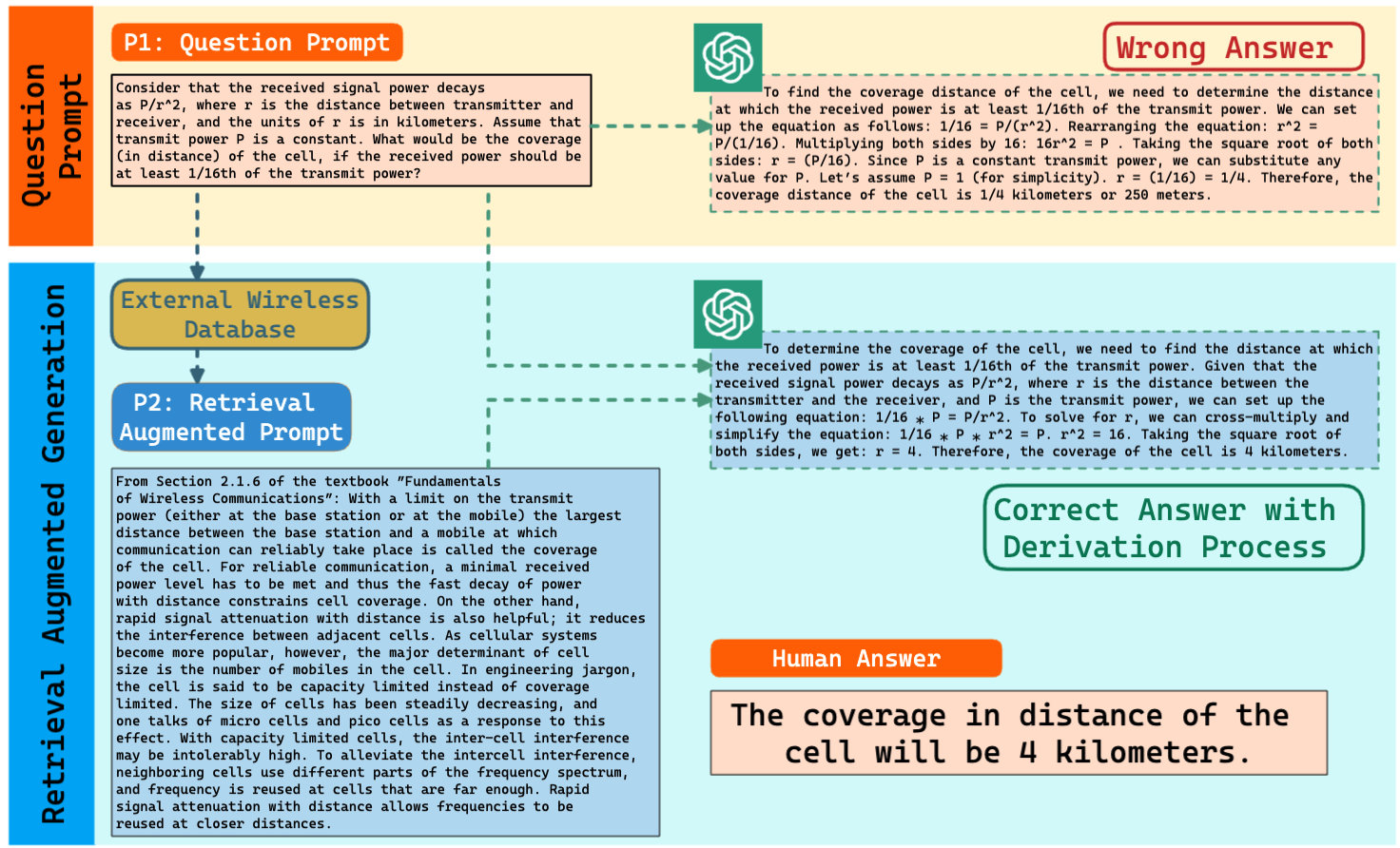}
}
\vspace{-1mm}
\caption{\small A sample mathematical Q/A pair from the dataset.}
\label{QA_1}\vspace{-0mm}
\vspace{-3.5mm}
\end{figure*}

\vspace{-1.5mm}\subsection{Instructiblity from Environmental Feedback}
\label{Instructibility}

For instructibility in LMMs, they should be able to dynamically adjust resource allocation, signaling policies, and many other cross-layer network functionalities in real time, catering to diverse tasks, environments, and optimization objectives. Additionally, they should be able to continuously monitor wireless observations to identify and address any unforeseen issues that might impede seamless network connectivity. A standard approach for dynamically adjusting wireless resource management and signaling schemes based on user feedback involves the use of deep RL. However, deep \ac{RL} techniques are task-specific and necessitate retraining when the wireless environment and optimization objectives change. While multi-task \ac{RL} solutions exist, they are mostly limited to specific domains or \ac{OSI} layers. Moreover, they lack the ability to continuously evolve their state and action space to keep pace with changes in standards or advancements in wireless technology. In addition to these limitations, conventional multi-task \ac{RL} solutions lack the capability to perform abductive reasoning, a crucial aspect for making inferences about missing data or determining the best explanations for the observed data. 
These qualities are essential for achieving dynamic adaptability and reconfigurability, which are necessary for resilience and intent management. They are also crucial for supporting abductive reasoning capabilities required for semantic communications and other related tasks. 
We next discuss the key components needed to provide instructibility to the LMM. We begin by detailing the framework incorporating communication context, prompting, and an online LMM with wireless environment feedback, contributing to establishing an instructible system. Subsequently, we explore how to gain crucial qualities, such as logical and mathematical reasoning, that are pivotal for constructing self-evolving and dynamically adaptable systems, thus achieving instructibility.

\subsubsection{Communication context}
Causal representations that are used to represent the physical symbols, similar to tokens in NLP-based LLMs, form the \emph{communication context} for an LMM. This context encapsulates critical aspects of the wireless communication scenario. The components included in the LMM context include:

\begin{itemize}
\item \textbf{Network setup} that includes
details about 1) \emph{communicating devices},
2) \emph{communication link} (downlink or uplink),  and 
3) \emph{physical topology} that describes the antenna configuration, as well as any miscellaneous network architecture.

\item \textbf{Communication constraints} including constraints on the total power and 
shared communication/computation resources aross frequency, time, and other dimensions.

\item \textbf{Wireless standards/text snippets} read using RAG (Section~\ref{RAG_fine-tuning}), that include excerpts from relevant wireless communication standards or documents, providing a contextual basis for the communication scenario. 

\item \textbf{End-to-end optimization objectives}
that may include quality-of-service (QoS) measures such as average throughput, delay, reliability/quality-of-experience. 

\item \textbf{Historical wireless data} that may involve diverse measurements such as uplink pilots, user feedback on channel quality indication, various sensing measurements, and received uplink signal measurements, among other relevant parameters.
\end{itemize}
Next, we explain the different methods to construct an online LMM by instructing it with the environmental feedback.

\subsubsection{Iterative prompting mechanism}
The iterative prompting mechanism is a process in which an LMM is guided  through multiple rounds of interaction with human prompts. In each iteration, the model receives feedback or additional information based on the generated output (e.g., the QoS results based on the  wireless policy provided by the LMM) from the previous round, and this feedback is used to refine and improve subsequent responses.

\subsubsection{Online LMM with wireless environment feedback using neuro-symbolic AI}
\label{onlineLMM}
To build autonomous next-generation wireless systems, eliminating human intervention, which is required in iterative prompting, is crucial. To address this limitation, we suggest building an online LMM framework. In this setup, LMM functions as the wireless policy and is operationally embedded within an interactive setting using online RL. This entails utilizing gathered wireless observations and feedback from the environment to iteratively enhance its functionality, aligning with goals expressed in wireless language. The formulation of the LMM-powered cross-layer network functionalities can be represented as a partially observable Markov decision process. Here, the states are defined by the communication context and prompts,  actions are represented by the wireless policy suggested by the LMM, and rewards are determined using performance metrics obtained from the wireless environment. If available, the network goal or intent can be articulated in natural language by the network operator.
To ensure continuous operation without disrupting connectivity, online LMMs should possess the ability to explain wireless observations and infer any missing data, necessitating logical reasoning capabilities. Furthermore, given that many wireless concepts can be expressed mathematically, LMMs must inevitably be capable of performing mathematical reasoning. This includes tasks such as channel predictions, beamforming vector computations, channel quality measurements, and many other cross-layer network computations.

In contrast to multi-task \ac{RL}, which lacks logical and mathematical reasoning capabilities, we advocate incorporating  those qualities through the use of  \emph{neuro-symbolic AI} \cite{ChristoTWCArxiv2022}. 
In our setting, symbolic AI serves to evaluate diverse logical and mathematical formulas, while the neural component is responsible for learning the logical and mathematical equations from wireless observations and context information. 
When prompted with communication context and grounded wireless observations using causal discovery (see Section~\ref{RAG_fine-tuning}), symbolic AI allows connecting facts and data through rules and intricate algorithms, resembling the cognitive operations of the human brain in storing high-level concepts and engaging in nuanced inferencing for appropriate responses. To prevent hallucination, LMMs must possess not only exceptional language interpretation and generation capabilities but also the ability to explain wireless observations and infer any missing data. Additionally, they should establish connections between various physical symbols and the rules acquired through symbolic AI. Here, a viable approach is to develop a formal logical language designed to encapsulate the exhaustive ontology of wireless concepts and articulate rules governing the functioning of wireless systems (and is the symbolic part here). This strategy is reminiscent of the Cyc concept \cite{LenatMarcusArxiv2023}, which serves a similar purpose for web-based data. Beyond their lack of logical reasoning abilities, existing LLMs face challenges in accurately capturing mathematical formulas and executing mathematical derivations. We propose to imbue LMMs with mathematical reasoning, empowering them to conduct computations specifically tailored for wireless systems. In this context, a promising approach involves leveraging a neuro-symbolic problem solver \cite{QinArxiv2023}, which comprises three main components. First, a \emph{problem reader} encodes math word problems, presented as textual prompts, into vector representations. Second, a \emph{programmer} generates symbolic grounded equations, which are executed to produce answers. Lastly, a \emph{symbolic executor} obtains final results. In this setup, the programmer entails learning the weights (neural part) that establish connections between various mathematical symbols. The resulting neuro-symbolic problem solver of enables the construction of dynamic problem solvers, a critical component for intent management and resilience, as discussed in Section~\ref{Intent_Management} and ~\ref{Resilience}.

\section{Experimental Validation: A Use Case Study for Intent Management and Resilient Networking}
\label{OpenProblems}
In this section, we first demonstrate illustrative experiments conducted on a dataset specific to wireless scenarios using RAG. We highlight the enhanced performance of LMM compared to vanilla LLMs which does not have any wireless context, characterized by succinct explanations (resulting in \emph{reduced hallucinations}), precise answers (demonstrating \emph{grounding in wireless concepts}), and well-founded rationales (illustrating \emph{mathematical reasoning} capabilities). Additionally, we discuss how the results indicate the potential application of the proposed LMM in addressing specific challenges in future wireless networks.

\subsection{Experimental Validation}
\label{Validation}

To evaluate the efficacy of RAG in wireless contexts, we conducted certain Q/A experiments, where a sample question from the dataset is shown in Fig.~\ref{QA_1}. 
In the RAG process, relevant paragraphs are extracted from \cite{TseCUP2005} to serve as wireless context information. This information encompasses a combination of textual content and mathematical symbols and equations. This structured wireless information represents a simple multi-modality case. Table ~\ref{tab:evaluation_metric} shows a human-involved evaluation (16 participants) on responses from different prompting methods of 4 conceptual wireless questions and 7 mathematical wireless questions. With the help of retrieved knowledge context, common LMM evaluation metrics, including precision, recall, F1 score, and ROUGE-L measure, indicate a performance improvement over vanilla LLMs ranging between $15\%$ to $30\%$. We also interpret the human evaluation result in the following way:
1) For conceptual questions, the standard \textit{Question Prompt} can retrieve reasonable rationale, while RAG can refine the assertion, leading to an $8\%$ improvement in the assertion metric. 
Furthermore, the over-explaining metric, which gauges the alignment of responses with human expectations, shows a remarkable almost three-fold improvement for RAG compared to vanilla LLMs.  
Improved assertions imply that LMMs can mitigate hallucinations in their responses, thereby aligning more effectively with the goals of the network.
2) 
In the context of mathematical questions, \textit{Retrieval Augmented Prompt} consistently provides the correct answer and can offer more detailed mathematical derivatives. Specifically, the rationale exhibits a $22\%$ improvement with RAG, while the derivative steps are $81\%$ more detailed than vanilla LLMs. This indicates that LMMs exhibit enhanced logical and mathematical reasoning abilities.

As discussed earlier, the demonstrated logical and mathematical reasoning capabilities, evidenced by improved rationale measures, enable LMMs to map wireless observations and context information fed as input to them into a mathematical problem formulation. This empowers LMMs to function as dynamic problem solvers (defined in Section~\ref{onlineLMM}). To exemplify the application of these capabilities in future wireless systems, we next delve into a few use cases, including intent management and resilience. 

\subsection{LMMs for Intent Management}
\label{Intent_Management}

A recent work that exploits LLMs for intent management appeared in \cite{TarkomaArxiv2023}. However, this prior work is limited to using an LLM as a chatbot to convert human specified intent in natural language to infrastructure level intents as network service descriptors. Furthermore, the authors in \cite{TarkomaArxiv2023} utilize human feedback to enhance the configurations generated by the LLM, hindering their ability to ensure intent assurance autonomously. We propose to go beyond this basic setup of \cite{TarkomaArxiv2023}. Specifically, we propose to employ LMMs for various phases in intent management spanned across the \ac{OSI} layers:
\begin{itemize}
    \item \emph{Problem formulation phase:} In intent-based wireless networks, the operator expresses business intents in simple natural language. The network autonomously translates these intents into an optimization problem, considering multiple objectives and physical constraints. 
An LMM can facilitate dynamic problem formulation without human intervention. This is achievable because, as validated using the metrics shown in Table.~\ref{tab:evaluation_metric}, LMMs possess better logical and mathematical reasoning abilities. Moreover, LMM responses are grounded in wireless physics, as discussed in Section~\ref{Grounding_Symbols}, and clarified in Fig.~\ref{QA_1}. It can thus formulate precisely wireless optimization problems, for resource allocation, signaling schemes, or a combination of cross-layer objectives.
    \item \emph{Intent assurance phase:} Leveraging a continuous stream of wireless measurements, LMMs can function as intent \emph{assurance agents}. These agents assess logical formulas representing the desired intent, typically defined by QoS targets. For evaluating the logical formulas, LMM utilize the neuro-symbolic AI capabilities of Section~\ref{Grounding_Symbols}. If the intent is not fulfilled, LMMs can identify and articulate the specific issues that need resolution, guiding efforts towards achieving intent assurance within a specified timeframe.
    \item \emph{Validator agent:} For solving the LMM-designed problem formulation provided we can use multiobjective \ac{RL} with causal reasoning games building on \cite{thomas2023causal}. Validating solutions against regulatory norms and physical constraints for long-term intent fulfillment is crucial. This is the \emph{alignment goal} described previously in Section I. To autonomously manage intent, a validator role can be fulfilled by LMM, possessing a solid understanding of wireless concepts.

\end{itemize}
Next, we delve into the impact of minimizing hallucinations through improved assertions and providing precise answers (as reflected in overexplaining metric). This capability is pivotal for swiftly recovering from network service disruptions and thereby ensuring resilience, especially when hallucinated responses tend to deviate off course. 
\begin{table}
    \centering
    \begin{tabular}{l|l|l}
        \toprule
        Evaluation Measure & \makecell{Question \\ Prompt} & \makecell{Retrieval \\ Augmented \\ Prompt}\\
        \midrule
        Precision ($\uparrow$) & 0.06 & \textbf{0.08} \\
        Recall ($\uparrow$) & 0.59 & \textbf{0.65} \\
        F1 Score ($\uparrow$) & 0.11 & \textbf{0.14} \\
        ROUGE-L (F-measure) ($\uparrow$) & 0.17 & \textbf{0.20} \\
        \midrule
        Over Explaining ($\downarrow$) & 0.34 & \textbf{0.12} \\
        Conceptual Question Rationale ($\uparrow$) & \textbf{0.89} & 0.84 \\
        Conceptual Question Assertion ($\uparrow$) & 0.88 & \textbf{0.95}\\
        Mathematical Question Rationale ($\uparrow$) & 0.77 & \textbf{0.94}\\
        Mathematical Question Assertion ($\uparrow$) & 0.76 & \textbf{0.97} \\
        Mathematical Question Derivative Steps ($\uparrow$) & 0.48 & \textbf{0.87} \\
        \bottomrule
    \end{tabular}
    \caption{\small Prompting with retrieval-augmented context shows a general advancement over purely prompting with questions in 4 quantitative measurements (upper section) and 6 human-evaluated measurements (lower section). For each metric (row), the symbol ($\uparrow$) indicates that higher scores are better, and better results are highlighted in \textbf{bold} for the two prompting methods. \textbf{Precision}: the number of shared words to the total number of words in the generated answers; \textbf{Recall}: the number of shared words to the total number of words in the human answers; \textbf{F1 score}: $\frac{2 \cdot Precision \cdot Recall}{Precision+Recall}$; \textbf{ROUGE-L (F-measure)}: based on the longest common subsequence (LCS) between the generated answer and human answer, which indicates that a longer shared sequence should indicate more similarity between the two sequences. \textbf{Human Evaluated Score}: Participants are asked to rate each Q/A sample without knowing the source of it. They will give score 0 for no and score 1 for yes for each of the Rationale, Assertion, and Over Explaining items. Mathematical questions require an additional derivative step to be scored.\vspace{-0.5cm}}
    \label{tab:evaluation_metric}
\end{table}
\subsection{LMMs for Resilient Networking}
\label{Resilience}

\emph{Resilience} can be defined as the ability of wireless networks to: a) to detect or predict in advance any failures or performance disruptions arising due to network functionality issues across any \ac{OSI} layer, changing wireless environment, user dynamics, or external malicious influences; and b) to recover back to their normal functionality within a stipulated time frame, thereby ensuring seamless connectivity for all connected devices.
 In \cite{thomas2023causal}, we proposed a robust framework for building resilient wireless networks using the principles of causal Bayesian optimization. However, the application of our solution in \cite{thomas2023causal} is limited, because it mainly focuses around quickly recovering from QoS deviations in the network. However, network service disruptions can stem from changes in the wireless environment or malfunctions in hardware or software functionalities across diverse edge devices. To address this challenge, we suggest leveraging LMMs equipped with causal knowledge, not only pertaining to the wireless environment but also grounded in wireless standards and cross-layer network functionalities. This approach involves using a universal foundation model that can handle service disruptions across multiple domains and tasks. We define resilient wireless systems using LMMs as composed of the following two stages that operates in a closed loop fashion as discussed next:
\begin{itemize}
    \item \emph{Continuous monitoring of service disruptions:} 
To detect network service disruptions, the LMM should continuously monitor and predict potential issues across \ac{OSI} layers. For example, consider a situation where the software code representing functionality at any \ac{OSI} layer on an edge device becomes corrupted due to processor malfunctions. Alternatively, critical information intended for storage on an edge server might face corruption due to jamming attempts. Herein, since LMMs are grounded in wireless concepts, they can adeptly analyze error messages and descriptions of software malfunctions or data corruptions, offering suggestions aligned with network standards to rectify the issues, without any human intervention. This approach enhances the model's capability to provide context-aware solutions across diverse tasks or domains or environments. Furthermore, LMMs can be consistently prompted to check for potential wireless environment issues that might lead to performance deviations in the near future. 
Such network issues can be formulated as either logical formulas or mathematical equations. For example, a potential logical formula could be $p: X \rightarrow (Y < \tau)$, signifying that with probability $p$, the wireless observation $X$ results in a performance $Y$ below the expected target $\tau$.  However, vanilla LLMs face challenges in handling such logical and mathematical problems, as discussed in Section~\ref{Instructibility}. In this context, logical and mathematical reasoning capabilities using neuro-symbolic problem solvers play a crucial role in assessing performance quality. 
In contrast to \cite{thomas2023causal}, which necessitates the construction of specific causal models for monitoring particular tasks or QoS targets, the universal nature of LMMs allows a single model to be used for monitoring performance deviations and software or hardware malfunctions across any \ac{OSI} layer. Given that a failure is detected, we next look at how the LMM can help the network functionalities quickly (within a stipulated time) recover back to the expected performance.
    \item \emph{Faster network service recovery via LMM:} 

As detailed in Section~\ref{RAG_fine-tuning}, RAG empowers LMM to extract relevant wireless information from literature and 3GPP standards. It enables the model to comprehend the causal implications of network actions by grounding wireless observations to the extracted knowledge, allowing it to execute the minimal interventions required to restore the network to a normal functioning state. Further, as discussed in Section~\ref{Instructibility}, instructibility employing iterative prompting or online LMMs with feedback from the wireless environment enables LMMs to generate a sequence of network actions. These actions can involve repairing malfunctioning code, adjusting resource allocation, or refining signaling schemes to restore the network to normal functioning. 
In contrast to the approach presented in \cite{thomas2023causal}, which might require separate causal AI models to monitor diverse functionalities across OSI layers, the universality of LMMs enables faster switching between tasks requiring repair or refinement, utilizing a single AI model.
\end{itemize}

  \section{Challenges and Open Questions for LMM-Empowered AI-Native Wireless Systems
}

The initial experiments utilizing RAG, as shown in Section~\ref{Validation}, offered valuable insights into LMM's capabilities when grounded with wireless concepts. However, specific challenges persist in our pursuit of meeting the future wireless network goals as discussed previously.

\subsection{How can We Enable LMMs to do Planning?}

Meeting future wireless network goals requires the capability of LMMs to provide recommendations across various \ac{OSI} layers. This includes resource allocation policies, waveforms under non-linear signal models, network slicing policies, and more. To ensure that these recommendations align with long-term network goals in terms of performance, sustainability, or seamless connectivity, LMMs must possess the ability to perform planning. In this context, planning entails the ability of LMMs to propose a sequence of multi-dimensional network actions, enabling the network to optimize performance objectives. The term ``multi-dimensional" reflects that these network actions are not restricted to a specific task or a single layer but can extend across multiple \ac{OSI} layers. Standard methods like deep \ac{RL} optimize actions for specific tasks but lack universality, as discussed in Section.~\ref{Vision}. Here, we look at possible approaches to incorporate planning in LMMs. The first approach, uses fine-tuning, and  involves taking a pretrained LLM and refining it using planning problems—consisting of instances and their solutions. While additional fine-tuning data and efforts might result in improved empirical performance, we must recognize that fine-tuning essentially transforms the planning task into a memory-based (approximate) retrieval process. This, however, falls short of providing conclusive evidence regarding the inherent planning capabilities of LMMs.

The second approach to enhance planning performance involves prompting the LMM with hints or suggestions to improve its initial plan guess. 
Crucial considerations in this context include whether the back prompting is manual or automated, the entity certifying the correctness of the final answer, and whether the prompts offer additional problem knowledge or merely encourage the LMM to reconsider its approach. A more popular methodology, here is ``chain of thought prompting (CoT)," that involves having a human (a system engineer) in the loop prompt the LMM. However, CoT is susceptible to the Clever Hans effect, where the LMM generates wireless policies, and the human in the loop, aware of right vs. wrong solutions, inadvertently guides the LMM. The responsibility for accuracy, if achieved, lies with the human in the loop. This framework raises concerns when the human cannot verify the answer to the planning problem themselves. However, in a wireless network, to ensure automated network operation, we cannot afford to rely on human intervention. Therefore, a promising approach entails the LMM critiquing its predictions through self-reflection capability and iteratively self-improving. This self-reflection capability can be achieved by incorporating causal reasoning. With the LMM being aware of each network action's causal effects, it can store these experiences in a short-term memory. Subsequently, when presented with new wireless observations, the LMM can self-reflect based on its previous experiences and engage in planning to ensure that network objectives are met.

\subsection{What are the Challenges Associated with Training a Universal Foundation Model?}

Training a foundation model for wireless communication enables domain-specific optimization, efficiency gains through reduced NN weights, and enhanced performance. However, achieving this long-term goal necessitates collaboration among stakeholders in wireless communication and computer science. The associated challenges include the need for seamless interdisciplinary cooperation, addressing diverse communication standards, and navigating evolving technologies to ensure the model's adaptability and effectiveness.
\begin{itemize}
    \item \emph{Diverse and representative datasets:} To ensure seamless connectivity across different wireless environments, diverse applications and services, the LMM should be trained under diverse signal conditions, interference patterns and fading scenarios.
    \item \emph{Adaptability to evolving standards and technologies:} The infusion of evolving 3GPP standards into the foundation models is crucial. This integration ensures that the decisions made by the LMM are in harmony with both the network and the unified 3GPP standards. This alignment contributes significantly to enhancing the trustworthiness of the language model by ensuring its compatibility and compliance with the latest industry standards. In addition to leveraging LMMs for understanding and adhering to existing standards, they can also play a crucial role in the creation of standards, especially in scenarios where technologies are not standardized. LMMs, with their capacity for natural language processing and generation grounded in wireless concepts, can contribute to the formulation and documentation of wireless standards, fostering innovation and clarity in technology development.

\end{itemize}

\subsection{How to Build Sustainable LMMs for Wireless Networks}

Employing pretrained LLMs like GPT-X or LLaMA with billions of parameters poses challenges for computational and memory-constrained wireless edge devices, limited by DRAM. Moreover, the vast energy consumption incurred by bulkier models during training and inference hinders the sustainability goals of future wireless systems. Possible solutions to overcome these challenges include:
\begin{itemize}
    \item \emph{Windowing} that allows to extract only relevant context information for a specific task, thereby reducing the computations involved in the attention mechanisms of LMMs. 
    \item Transformer NN weights show \emph{sparse structure} as highlighted in Falcon LLM \cite{AlmazroueiDebbahArxiv2023}. Employing \emph{sparse inference} techniques becomes advantageous in this context as it aids in diminishing computational complexity by eliminating all the insignificant weights.
    \item A universal foundation model can be used as an anchor model, residing in the cloud. Employing knowledge distillation, this anchor model imparts knowledge to edge servers, guiding them in creating a specialized \emph{short language model} designed for the precise wireless tasks required. This distributed architecture optimizes NN parameter usage and helps reduce the computational complexity during inference.
\end{itemize}

\section{Conclusion and Recommendations}
This article developed a new framework for designing AI-native wireless systems (6G and beyond) for multiple tasks using foundation models built on multi-modality, grounding, and instructibility principles. We conclude with three key
recommendations:
\begin{itemize}
    \item \textbf{Speeding up next-G standardization to system design:} LMMs can assist in the swift prototyping of diverse system design scenarios. Leveraging the capabilities of RAG, LMMs can retrieve pertinent text-based descriptions and specifications by considering the provided input, whether it be a network intent or system design goals. This enables LMMs to actively contribute to rapidly exploring design alternatives and their associated implications.
    \item \textbf{Building a repository of wireless datasets:} The results based on RAG that we presented offer unique insights into the capabilities of LMMs grounded in wireless concepts. However, it is essential to acknowledge the challenges associated with generating a comprehensive dataset. To address this, we recommend the creation of an open-source ontology for wireless concepts and algorithms, sourced from a curated selection of textbooks and wireless literature, encompassing 3GPP standards. This approach ensures the quality, reliability, and trustworthiness of the dataset, making it applicable for research and development across the entire wireless community.
    \item \textbf{Compositions of short language models and distributed architecture:} 
Creating a universal foundation model at each base station may prove impractical due to the substantial energy consumption and computational resources involved. In this context, we recommend a distributed architecture that involves constructing distinct short language models learned at edge servers.
These edge servers might require only a condensed language model, as the applications or tasks they handle are limited in scope. Furthermore, such a distributed architecture facilitates collaborative reasoning based on the principle of compositionality, as outlined in \cite{BansalArxiv2024}, through the combination of representations from multiple smaller models. This compositional approach empowers the LMM-based network with the capacity to acquire diverse skills and functionalities over time, and thus gaining universal capabilities in a distributed fashion.
\end{itemize}
\bibliographystyle{IEEEbib}
\bibliography{bibliography,semantics_refs,bib,LLM_refs}

\end{document}